%
\let\useblackboard=\iftrue
%
%
\newfam\black
\input harvmac.tex
\input tables.tex
%
\input epsf.tex
\ifx\epsfbox\UnDeFiNeD\message{(NO epsf.tex, FIGURES WILL BE
IGNORED)}
\def\figin#1{\vskip2in}
\else\message{(FIGURES WILL BE INCLUDED)}\def\figin#1{#1}\fi
\def\ifig#1#2#3{\xdef#1{fig.~\the\figno}
\midinsert{\centerline{\figin{#3}}%
\smallskip\centerline{\vbox{\baselineskip12pt
\advance\hsize by -1truein\noindent{\bf Fig.~\the\figno:} #2}}
\bigskip}\endinsert\global\advance\figno by1}
\noblackbox
\def\Title#1#2{\rightline{#1}
\ifx\answ\bigans\nopagenumbers\pageno0\vskip1in%
\baselineskip 15pt plus 1pt minus 1pt
\else
\def\listrefs{\footatend\vskip
1in\immediate\closeout\rfile\writestoppt
\baselineskip=14pt\centerline{{\bf
References}}\bigskip{\frenchspacing%
\parindent=20pt\escapechar=` \input
refs.tmp\vfill\eject}\nonfrenchspacing}
\pageno1\vskip.8in\fi \centerline{\titlefont #2}\vskip .5in}
 
scaled\magstep3
 
scaled\magstep3
 
scaled\magstep3
 
scaled\magstep3
 
scaled\magstep3
\ifx\answ\bigans\def\tcbreak#1{}\else\def\tcbreak#1{\cr&{#1}}\fi
\useblackboard
\message{If you do not have msbm (blackboard bold) fonts,}
\message{change the option at the top of the tex file.}

\font\blackboard=msbm10 scaled \magstep1
\font\blackboards=msbm7
\font\blackboardss=msbm5
\textfont\black=\blackboard
\scriptfont\black=\blackboards
\scriptscriptfont\black=\blackboardss

\else

\fi
%
\def\ncr{ {\bf N_c}}
\def\oner{ {\bf 1}}
\def\mfr{{\bf m_f}}
\def\evr{{\bf 8_v}}
\def\yboxit#1#2{\vbox{\hrule height #1 \hbox{\vrule width #1
\vbox{#2}\vrule width #1 }\hrule height #1 }}
\def\fillbox#1{\hbox to #1{\vbox to #1{\vfil}\hfil}}
\def\ybox{{\lower 1.3pt \yboxit{0.4pt}{\fillbox{8pt}}\hskip-0.2pt}}

\def\comments#1{}

\def\half{{1\over 2}}
\def\Tr{{{\rm Tr\  }}}

\def\CN{{\cal N}}

\def\II{\relax{I\kern-.07em I}}
\def\IIA{{\II}A}

\def\inbar{\,\vrule height1.5ex width.4pt depth0pt}
\def\IZ{\relax\ifmmode\mathchoice
{\hbox{\cmss Z\kern-.4em Z}}{\hbox{\cmss Z\kern-.4em Z}}
{\lower.9pt\hbox{\cmsss Z\kern-.4em Z}}
{\lower1.2pt\hbox{\cmsss Z\kern-.4em Z}}\else{\cmss Z\kern-.4em
Z}\fi}
\def\IB{\relax{\rm I\kern-.18em B}}
\def\IC{{\relax\hbox{$\inbar\kern-.3em{\rm C}$}}}
\def\ID{\relax{\rm I\kern-.18em D}}
\def\IE{\relax{\rm I\kern-.18em E}}
\def\IF{\relax{\rm I\kern-.18em F}}
\def\IG{\relax\hbox{$\inbar\kern-.3em{\rm G}$}}
\def\IGa{\relax\hbox{${\rm I}\kern-.18em\Gamma$}}
\def\IH{\relax{\rm I\kern-.18em H}}
\def\IK{\relax{\rm I\kern-.18em K}}
\def\IP{\relax{\rm I\kern-.18em P}}
\def\pp{{\relax{=\kern-.42em |\kern+.2em}}}

\font\cmss=cmss10 \font\cmsss=cmss10 at 7pt
\def\IR{\relax{\rm I\kern-.18em R}}

\def\Tr{{\rm Tr\ }}


%
%

\def\NP{{\it Nucl. Phys.\ }}

\def\PL{{\it Phys. Lett.\ }}

\def\PRL{{\it Phys. Rev. Lett.\ }}

\Title{ \vbox{\baselineskip12pt\hbox{hep-th/9801002}
\hbox{BROWN-HET-1107}\hbox{TUW-97-19}
}}
{\vbox{
\centerline{Duality of Chiral $\CN=1$ Supersymmetric}
\centerline{Gauge Theories via Branes}}}

\centerline{Karl Landsteiner$^\natural$, Esperanza Lopez$^\natural$ and David
A.
Lowe$^\flat$}
\medskip
\centerline{$^\natural$ Institut f\"ur theoretische Physik, TU-Wien}
\centerline{Wiedner Hauptstra{\ss}e 8-10}
\centerline{A-1040 Wien, Austria}
\centerline{\tt landstei@tph45.tuwien.ac.at}
\centerline{\tt elopez@tph16.tuwien.ac.at}
\medskip
\centerline{$^\flat$Department of Physics}
\centerline{Brown University}
\centerline{Providence, RI 02912, USA}
\centerline{\tt lowe@het.brown.edu}
\bigskip

Using a six-orientifold on top of a NS-fivebrane we construct
a chiral $\CN=1$ supersymmetric gauge theory in four dimensions with
gauge
group $SU(N_c)$ and
matter in the symmetric, antisymmetric and (anti)fundamental representations.
Anomaly cancellation is fulfilled by the requirement of a smooth
RR $7$-form charge distribution and leads us to the introduction
of $8$ half D-sixbranes ending on the NS-fivebrane. We obtain the
dual model
from branes by a linking number argument. We check explicitly the
't Hooft anomaly matching conditions and the map between deformations
in the original and the dual model.

\Date{December, 1997}

\lref\intril{K. Intriligator, R.G. Leigh and M.J. Strassler, ``New
Examples of Duality in Chiral and Non-Chiral Supersymmetric Gauge
Theories,'' \NP {\bf B456} (1995) 567, hep-th/9506148.}
\lref\leigh{R.G. Leigh and M.J. Strassler, ``Duality of $Sp(2N_c)$
and
$SO(N_c)$ Supersymmetric Gauge Theories with Adjoint Matter,''
hep-th/9505088.}
\lref\seiberg{N. Seiberg, ``Electric-Magnetic Duality in
Supersymmetric Non-Abelian Gauge Theories,'' \NP {\bf B435} (1995) 129,
hep-th/9411149.}
\lref\brodie{J.H. Brodie and A. Hanany, ``Type IIA Superstrings,
Chiral Symmetry and $\CN=1$ 4D Gauge Theory Dualities,'' \NP {\bf B506}
(1997) 157, hep-th/9704043.}
\lref\clifford{N. Evans, C.V. Johnson and A. Shapere, ``Orientifolds,
Branes and Duality of 4D Gauge Theories,'' \NP {\bf B505} (1997) 251,
 hep-th/9703210.}
\lref\lll{K. Landsteiner, E. Lopez and D.A. Lowe, ``$\CN=2$
Supersymmetric Gauge Theories, Branes and Orientifolds,'' \NP {\bf B507}
(1997) 197,
hep-th/9705199.}
\lref\barbon{J.L.F. Barbon, ``Rotated Branes and $\CN=1$ Duality,''
\PL {\bf B402} (1997), 59, hep-th/9703051.}
\lref\kutasov{S. Elitzur, A. Giveon and D. Kutasov, ``Branes and
$\CN=1$ Duality in String Theory,''  \PL {\bf B400} (1997) 269,
hep-th/9702014,
S. Elitzur, A. Giveon, D. Kutasov, E. Rabinovici and A. Schwimmer,
``Brane Dynamics and $\CN=1$ Supersymmetric Gauge Theory,'' \NP B505 (1997)
202, hep-th/9704104.}
\lref\hanwit{A. Hanany and E. Witten, ``Type IIB Superstrings, BPS
Monopoles, and Three-Dimensional Gauge Dynamics,'' \NP {\bf B492} (1997) 152,
hep-th/9611230.}
\lref\hazaff{A. Hanany and A. Zaffaroni, ``Chiral Symmetry from Type
IIA Branes,'' hep-th/9706047.}
\lref\lpt{J. Lykken, E. Poppitz and S. P. Trivedi, ``Chiral Gauge
Theories
from D-Branes,'' hep-th/9708134; ``M(ore) on Chiral Gauge Theories
from
D-Branes,'' hep-th/9712193.}
\lref\supered{E. Witten, ``Solutions of four-dimensional Field
Theories via M-Theory,'' \NP {\bf B500} (1997), 3, hep-th/9703166.}
\lref\joe{ J. Polchinski, ``Dirichlet Branes and Ramond-Ramond
Charges,''
\PRL {\bf 75} (1995) 4724, hep-th/9510017
and
``TASI Lectures on D-branes,'' preprint NSF-ITP-96-145
hep-th/9611050.}
\lref\karlesp{K. Landsteiner and E. Lopez, ``New Curves from
Branes,''
hep-th/9708118.}
\lref\future{K. Landsteiner, E. Lopez and D.A. Lowe, work in
progress.}
\lref\ooguri{K. Hori, H. Ooguri and Y. Oz, ``Strong Coupling Dynamics
of Four-Dimensional $\CN=1$ Gauge Theories from M Theory Fivebrane,''
hep-th/9706082.}
\lref\andi{A. Strominger, ``Open p-Branes,'' \PL {\bf B383} (1996) 44;
P. Townsend, ``D-Branes from M-Branes,'' \PL {\bf B373} (1996) 68,
hep-th/9512062.}
\lref\vectorwit{E. Witten, ``Toroidal Compactification Without Vector
Structure,'' hep-th/9712028.}
\lref\seiprobe{N. Seiberg, ``IR Dynamics on Branes and Space-Time
Geometry,'' \PL {\bf B384} (1996) 81, hep-th/9606017.}
\lref\threesw{N. Seiberg and E. Witten, ``Gauge Dynamics And Compactification
To Three Dimensions,'' hep-th/9607163.}
\lref\orientis{A. Brandhuber, J. Sonnenschein, S. Theisen and S.
Yankielowicz,
``M Theory And Seiberg-Witten Curves: Orthogonal and Symplectic Groups,''
\NP {\bf B504} (1997) 175, hep-th/9705232;
C. Ahn, K. Oh and R. Tatar, ``Branes, Geometry and $\CN=1$ Duality with
Product Gauge Groups of SO and Sp,'' hep-th/9707027;
C. Csaki and W. Skiba, ``Duality in Sp and SO Gauge Groups from M
Theory,''
hep-th/9708082;
C. Ahn, K. Oh and R. Tatar, ``$Sp(N_c)$ Gauge Theories and M Theory
Fivebrane,''
hep-th/9708127;
C. Ahn, K. Oh and R. Tatar, ``M Theory Fivebrane Interpretation for Strong
Coupling Dynamics of $SO(N_c)$ Gauge Theories,'' hep-th/9709096;
C. Ahn, K. Oh and R.  Tatar, ``M Theory Fivebrane and Confining Phase of
$\CN=1$ $SO(N_c)$ Gauge Theories,'' hep-th/9712005;
C. Ahn, ``Confining Phase of $\CN=1$ $Sp(N_c)$ Gauge Theories via M Theory
Fivebrane,'' hep-th/9712149;
S. Terashima, ``Supersymmetric Gauge Theories with Classical Groups via M
Theory Fivebrane,'' hep-th/9712172.}

\newsec{Introduction}

In the last year Dirichlet brane \joe\ techniques have been proven to
be
an
extremely
powerful tool for the study of non-perturbative phenomena in
supersymmetric
gauge theories. A collection of parallel D-p-branes would give gauge
theories with 16 supercharges.
In order to reduce the number of supersymmetries one can use the
fact that D-branes can end on certain other branes \andi. In Type \IIA\
string theory
D-fourbranes are allowed to end on Neveu-Schwarz (NS)-fivebranes.
Connecting parallel NS-fivebranes by
a set of D-fourbranes leads then to an effectively four-dimensional
$\CN=2$ gauge
theory \supered.
Matter hypermultiplets transforming in the fundamental representation
can be introduced by including D-sixbranes.
When a D-sixbrane crosses one of the NS-fivebranes
a D-fourbrane stretched between the NS-fivebrane and the D-sixbrane is created
\hanwit\ and this
leads to an alternative way of introducing hypermultiplets, i.e. one
gets
a hypermultiplet from strings stretching between fourbranes that end
on
different sides of a NS-fivebrane. Configurations with more than two
NS-fivebranes
lead to theories with product gauge groups and hypermultiplets
transforming in the bifundamental representation of two neighboring
gauge groups. Furthermore it has recently been shown that by putting
an orientifold sixplane on top of a NS-fivebrane one gets
hypermultiplets
in the symmetric or antisymmetric representation of $SU(N_c)$
\karlesp.
These brane configurations can be lifted to M-theory. The D-fourbranes
are then M-fivebranes wrapped around the eleventh dimension. In this
limit one
deals with a single smooth M-fivebrane wrapped on a complex one
dimensional curve.
The complex curves are precisely the ones which parameterize the
Coulomb-branch
of $\CN=2$ gauge theories.

By rotating the NS-fivebranes with respect to each other one can further
reduce
the amount of supersymmetry. Choosing in particular angles that
correspond
to $SU(2)$ rotations in a four-dimensional embedding space one can
construct
brane configurations with $\CN=1$ supersymmetry in four dimensions
\barbon.
It was
shown in \kutasov\ that Seiberg's dualities \seiberg\ can be recovered by
moving on the
moduli space of the brane configuration. An important feature that is
present
for $\CN=1$ but not for $\CN=2$ is chirality. A mechanism for
enhanced chiral
symmetry in the brane configuration has been proposed in \brodie.
There it
was suggested that when $N_f$ D-sixbranes meet a NS-fivebrane such that
they are
divided in two pieces one gets an enhanced $SU(N_f)_R \times
SU(N_f)_L$.
This leads to the picture that strings stretching from half D-sixbranes
ending on NS-fivebranes to the D-fourbranes give rise to chiral $\CN=1$
multiplets.
The fact that the Ramond-Ramond (RR) 7-form charge is not conserved for such a
configuration
is interpreted as the anomaly in the $\CN=1$ gauge theory \hazaff.
In order to
generate anomaly free chiral models with a simple gauge group one
needs
matter in two index tensor representations\foot{A different way of
obtaining anomaly free chiral models from
branes has been proposed in \lpt. Here one puts the brane
configuration
in an orbifold background. The chiral matter is produced by the
orbifold
projection rather than by half D-sixbranes.}. Thus one would expect
that by
combining the methods of \hazaff\ with that of \karlesp\ one should be
able
to construct brane configurations
realizing anomaly free $\CN=1$ gauge theories. We show that this
is indeed true.

\newsec{Brane Configurations}

Let us begin by reviewing the basic brane configuration we will use
to realize four-dimensional gauge theories with $\CN=1$
supersymmetry.
The system of branes includes three different NS-fivebranes.
The central fivebrane (labeled $B$) extends in the
$(x^0,x^1,x^2,x^3,x^8,x^9)$ directions, while the outer fivebranes
(labeled $A$ and $C$) are tilted at $SU(2)$ angles from the $(x^8,x^9)$ plane
towards the $(x^4,x^5)$ plane.
We also have a number of D-fourbranes stretching between
the fivebranes extended along the $(x^0,x^1,x^2,x^3,x^6)$ directions.
Finally we have three different types of D-sixbranes, which stretch
parallel to the three different kinds of fivebranes, but are also
extended in the $x^7$ direction. This configuration of branes gives
rise to a gauge theory with $\CN=1$ supersymmetry in the four
dimensions $(x^0,x^1,x^2,x^3)$, for arbitrary angles.

The gauge group and matter content are determined by the following
rules.

1) $n$ D-fourbranes stretching between two fivebranes will give rise
to a massless vector multiplet with $SU(n)$ gauge symmetry. When the
fivebranes are parallel there will be an additional chiral multiplet
in the adjoint representation, however this multiplet becomes massive
when the branes are at an angle.

2) A D-fourbrane between two D-sixbranes gives  rise to two chiral
multiplets.

3) When a D-fourbrane stretches between a D-sixbrane and a parallel
NS-fivebrane, there is a single chiral multiplet. If the fivebrane is
not parallel to the sixbrane, no massless matter appears.

4) D-fourbranes on one side of an NS-fivebrane give rise to a chiral
multiplet in the fundamental and one in the anti-fundamental
representation of the gauge group associated with fourbranes on the
other side of the NS-fivebrane.

5) When a D-sixbrane meets a D-fourbrane in space a chiral multiplet
in the fundamental and one in the anti-fundamental representation is
generated. As argued by Brodie and Hanany \brodie, when the sixbranes
touch a parallel fivebrane, chiral symmetry is restored, and the
sixbranes may split.

\ifig\electric{A configuration of three fivebranes connected by $N_c$
parallel
fourbranes and an orientifold sixplane. The outer fivebranes are
rotated
by an angle $\vartheta$ with respect to the middle one. The O6 is
divided in
half by the middle fivebrane in the $x^7$-direction and changes its
character
there. To compensate for the jump in the sixbrane charge of the
orientifold
we also have to put
eight half D-sixbranes on top of the orientifold where it has charge
$-4$.
These sixbranes end on the middle fivebrane. There are also $m_f$
additional
sixbranes (and their mirror images) tilted in the same way as the
fivebranes.
They also extend along
the $x^7$-direction (this is not shown in the figure).}
{
\epsfxsize=5truein\epsfysize=4truein
\epsfbox{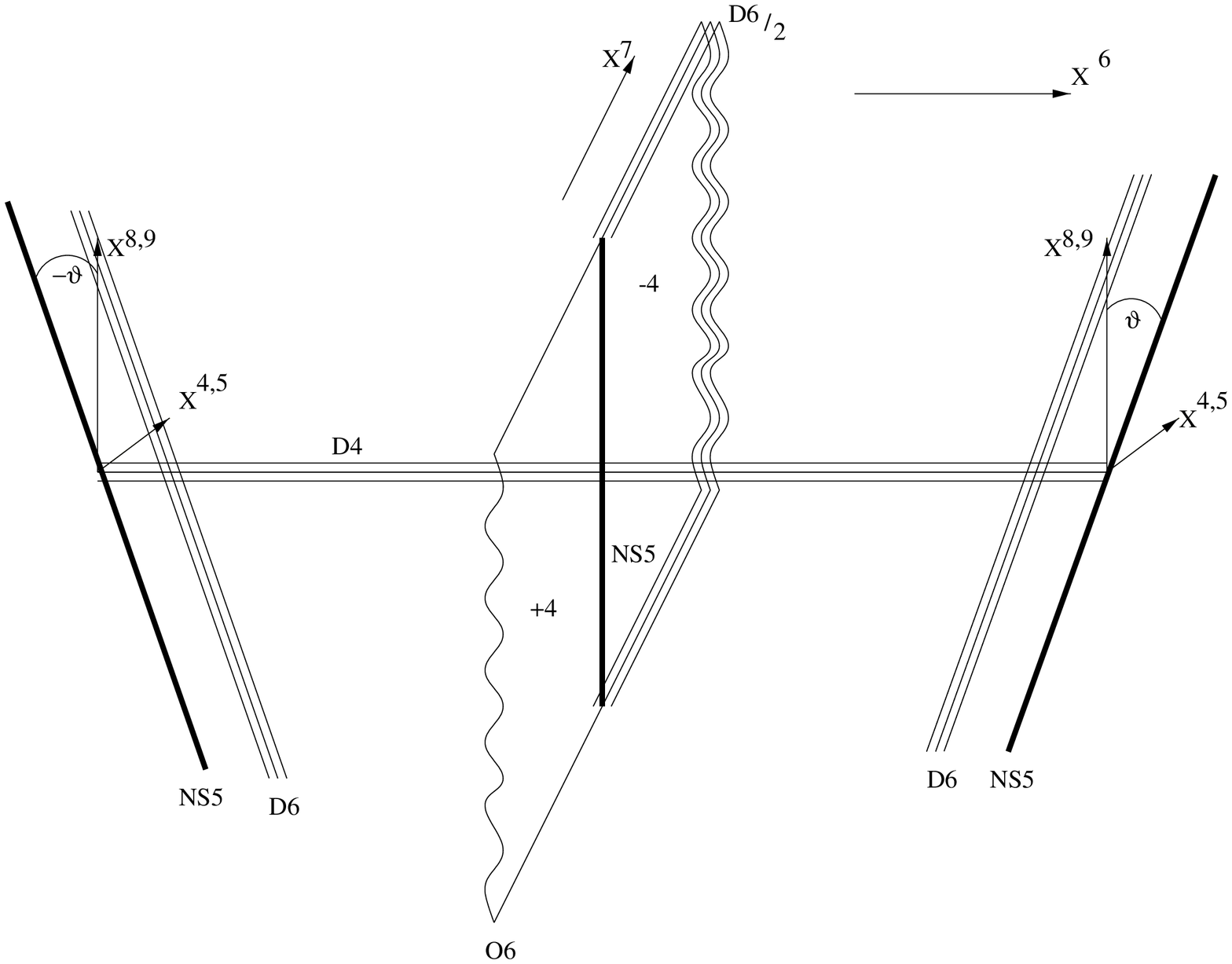}}

To construct a gauge theory with truly chiral matter we will use one
more element, an orientifold sixplane, which we will place on the
central fivebrane, extending in the $(x^0,x^1,x^2,x^3,x^7,x^8,x^9)$
directions. The brane configuration is shown in \electric. The
orientifold acts as a reflection taking
$(x^4,x^5,x^6) \to (-x^4,-x^5,-x^6)$.
{}From now on all branes will be arranged to respect
this $\IZ_2$ symmetry.

Orientifolds have been used in previous works
\refs{\clifford,\lll,\karlesp,\orientis} to construct gauge theories with
orthogonal and symplectic gauge groups using brane constructions.
There are two possible choices of orientifold projection, one of
which gives rise to orthogonal gauge groups on the D-fourbranes with a
symplectic symmetry appearing on D-sixbranes. For the case of an
orientifold sixplane, the sixplane carries RR charge
equivalent to 4 sixbranes. It has been argued that this orientifold
is equivalent to a $D_4$ singularity from the eleven-dimensional
M-theory point of view \refs{\karlesp, \vectorwit}.
The other choice of projection
gives rise to
symplectic groups on the D-fourbranes, and orthogonal groups on the
D-sixbranes. This orientifold sixplane carries $-4$ RR charge, and
has been argued to be equivalent to the Atiyah-Hitchin space from the
M-theory point of view \refs{\seiprobe, \threesw}.

In \clifford, it was argued that when an orientifold fourplane traverses
a NS-fivebrane the type of projection changes. In our case an orientifold
sixplane crosses the NS-fivebrane. This is T-dual to
the situation considered in \clifford.
Thus the orientifold projection must
change when the sixplane crosses a NS-fivebrane. In the case at
hand, this means the projection will change as one moves from
positive $x^7$ to negative $x^7$. We also place 8 half D-sixbranes on
the fivebrane to cancel the discontinuity in the RR charge. This will
lead to a smooth brane configuration from the eleven-dimensional
point of view, which will allow a geometric analysis of the $\CN=1$
duality in the spirit of \kutasov.

The gauge group of this configuration may be determined following
\karlesp\ to be $SU(N_c)$ where $N_c$ is the number of D-fourbranes.
Similar arguments imply the matter content is chiral with a chiral
multiplet in the symmetric representation, a chiral multiplet in the
antisymmetric representation, and $m_f+8$ chiral
multiplets in the fundamental, and $m_f$ chiral
multiplets in the anti-fundamental. The matter content is the same as
one of the models considered in \intril. However in our situation the
global symmetries and the superpotential are different. The two
classes of models are related by deformations, as we will see later.

Let us consider the superpotential that arises from this brane
configuration. We will argue it takes the following form
\eqn\superpot{
W = {1\over \mu} \Tr (X \tilde X)^2 + \hat Q \tilde X \hat Q~,
}
where $X$ is the antisymmetric field, $\tilde X$ is the symmetric
field and $\hat Q$ are the 8 fundamental fields arising from the 8
half D-sixbranes. The $(X \tilde X)^2$ term is seen to arise as in
\refs{\barbon,\brodie}. In our case, the two outer fivebranes $A$ and
$C$ are
rotated by equal but opposite angles, with respect to $B$. When the
angle is zero we have  an additional
massless chiral multiplet $\Phi$ in the adjoint representation.
There is a coupling of the form $\tilde X \Phi X$, but no coupling
of $\Phi$ to the fields $Q$, $\hat Q$ and $\tilde Q$. This is because
the sixbranes are parallel to the fivebranes.
For finite
angle, this chiral multiplet gets a mass given by $\tan \vartheta$.
Integrating out this field
gives the $(X \tilde X)^2$ term, with $\mu$ equal to the mass of the
adjoint chiral multiplet.

The 8 half D-sixbranes give rise to eight chiral multiplets in the
fundamental representation. Taking also the $m_f$ chiral multiplets
from the other D-sixbranes one would naively expect an $SU(m_f+8)_L$ flavor
symmetry. The flavor symmetry should appear as the gauge group on
the sixbranes. This is however not what the brane configuration tells
us. The 8 half D-sixbranes sit on top of an orientifold.
Therefore the flavor symmetry is $SU(m_f)_L\times SO(8)_L$.
The  $\hat Q \tilde X \hat Q$ term in the superpotential gives us
precisely this breaking of the flavor symmetry.
The origin of the $\hat Q \tilde X \hat Q$ term can also be seen when one
considers detaching
the fourbranes from the central fivebrane, and then moving the
fivebrane in the positive $x^7$ direction. The $\hat Q$ fermions
clearly become massive under this deformation. The gauge group is
broken to $SO(N_c)$ because now the fourbranes will cross the
orientifold sixplane in the region which gives a projection to the
orthogonal gauge group.
The position of the fivebrane in the $x^7$ direction is a single real
parameter, and plays the same role as the Fayet-Iliopoulos (FI)
parameters that appear in \kutasov. While the $U(1)$ component of the
gauge group is frozen, the FI parameters still appear in the D-term
equations of the low-energy field theory
\eqn\dterms{
\nu \delta_a^b = Q_a^f (Q^{\dag})^b_f + {\hat Q}_a^i ({\hat
Q}^{\dag})^b_i-  ({\tilde Q}^{\dag})_a^{\dot g}{\tilde Q}^b_{\dot g}
+
2 X_{ac} (X^{\dag})^{cb} - 2 (\tilde X^{\dag})_{ac} \tilde X^{cb}~,
}
where $\nu$ is related to the $x^7$ displacement of the fivebrane,
and
we have also introduced the fundamental fields $Q$ and the
anti-fundamentals $\tilde Q$.
We see then that taking $\nu$ to be negative (i.e. moving in the
positive $x^7$ direction), and assuming the fundamental,
anti-fundamental and antisymmetric squarks do not get a vev,
$\tilde X$ gets a vev. In the
simplest case where only the central fivebrane is moved, we expect
this to correspond to a baryonic branch characterized by
$(\tilde X)^{N_c}$ receiving a vev. This vev generates a mass for the
$\hat Q$'s provided the $\hat Q \tilde X \hat Q$ term appears in the
superpotential. Another way to see that the $\hat Q \tilde X \hat Q$
appears in the superpotential is to consider moving the central
fivebrane in the negative $x^7$ direction, which corresponds to an
expectation value for the
baryon $X^{N_c/2}$, for $N_c$ even.  The $\hat Q$ fermions will
remain massless in this case, and the gauge group will be broken to
$Sp(N_c)$ since now the fourbranes cross the orientifold plane on the
side that gives the symplectic projection. The $\tilde X$ field
becomes the adjoint of $Sp(N_c)$ with finite mass dependent on the
angle $\vartheta$. If we set $m_f=0$ and rotate to $\vartheta=\pi/2$,
$\tilde X$
will be massless and $\CN=2$ supersymmetry will be restored, which
requires a coupling of the form $\hat Q \tilde X \hat Q$.
We will postpone further discussion of these baryonic
branches to the next section.

\ifig\magnetic{The magnetic brane configuration}
{
\epsfxsize=4.5truein\epsfysize=4truein
\epsfbox{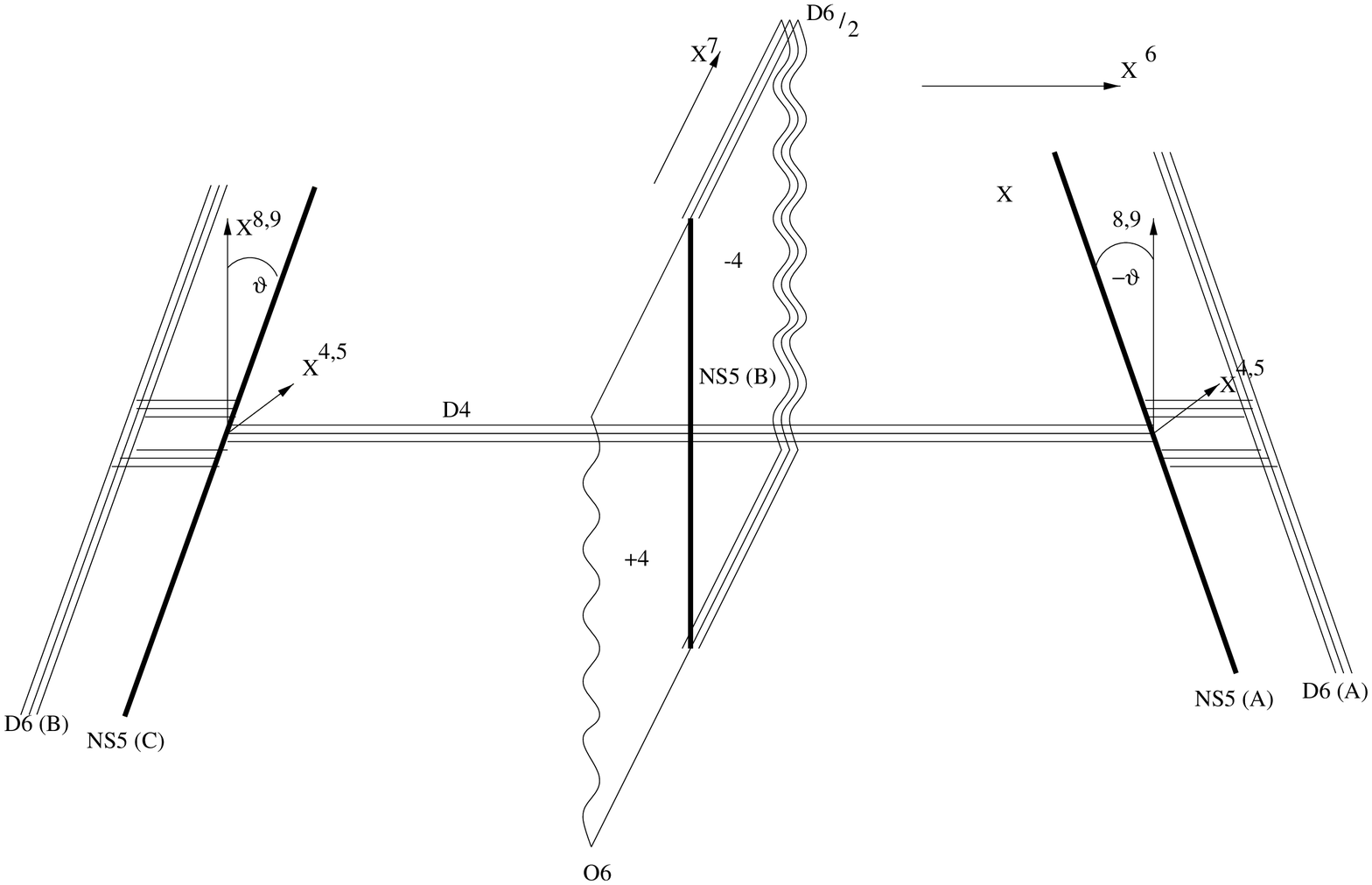}}

{}From the brane point of view, the transition to the magnetic dual
theory is seen by moving the branes to a mirror image configuration,
and then moving the sixbranes outside the fivebranes.
This is shown in figure \magnetic. Since for our brane configuration
we do
not expect the orientifolding to introduce any extra singularities,
we may use the linking number argument
\hanwit\ to determine how many D-fourbranes we end up with after
rearranging the fivebranes and sixbranes. We assume a fourbrane is
created whenever a fivebrane crosses a sixbrane which is non-parallel
\brodie.
In addition, we assume that when a fivebrane crosses the orientifold
plane,
4 D-fourbranes are created (i.e. the orientifold plane is treated as
set of 4 D-sixbranes as far as the linking number argument goes).
The linking number of a fivebrane is given by
$L_{5}= \half( n_{6L} -n_{6R})+(n_{4R} - n_{4L})$, where $n_{6L}$
is the
number of non-parallel D-sixbranes to the left, etc.
Similarly the linking number
of a D-sixbrane is given by $L_{6} = \half( n_{5L} -n_{5R})+(n_{4R}
-
n_{4L})$
where $n_{5L,R}$ denotes now the number of non-parallel fivebranes to
the
left
or to the right of the D-sixbrane under consideration.
These numbers are conserved under
brane moves. If we focus on a single $A$ sixbrane
and
the fivebrane $A$, their initial linking numbers
are $L^A_{6} = -1$ and
$L^A_{5} = - \half m_f -2 +N_c $. In the final configuration,
the
linking
number of the D-sixbrane is ${L'}^A_{6} = 1 - n_4$, where $n_4$ is
the
number
of D-fourbranes connecting an A-D-sixbrane with the fivebrane A.
Conservation
of the linking number gives $n_4=2$. The linking number of the
fivebrane
is
${L'}^A_{5} = 2+{5\over 2} m_f - \tilde N_c$.
With this we find
$\tilde N_c = 3 m_f - N_c +4$, and thus the dual gauge group is
$SU(\tilde N_c)$ with chiral multiplets in the symmetric and
antisymmetric representations, and $m_f+8$ fundamental chiral
multiplets, and $m_f$ anti-fundamental chiral multiplets. In
addition,
we have a number of gauge singlets corresponding to meson fields.

\newsec{Duality of Chiral Models}

\subsec{Electric Theory}

The gauge group is $SU(N_c)$ with a symmetric and an antisymmetric
chiral multiplet
together with $m_f+8$ fundamental and $m_f$ anti-fundamental
chiral multiplets. The anomaly-free global symmetry group is $SU(m_f)_L \times
SU(m_f)_R \times SO(8)_L \times U(1)_B \times U(1)_X \times U(1)_R$.
The fields live in the following representations:
\thicksize=1pt
\vskip12pt
\begintable
\tstrut  | $SU(N_c)$ | $SU(m_f)_L$ | $SU(m_f)_R$ | $SO(8)_L$
| $U(1)_B$ | $U(1)_X$ | $U(1)_R$ \crthick
$\hat Q$ | $\ncr$ | $\oner$ | $\oner$ | $\evr$ | ${1\over N_c}$ | $\half$
|
${3\over 4}$ \cr
$Q$ | $\ncr$ | $\mfr$ | $\oner$ | $\oner$ | ${1\over N_c}$ | $0$ |
$1-{{N_c-2}\over{2 m_f}}$ \cr
$\tilde Q$ | $\overline \ncr$ | $\oner$ |  $\mfr$ | $\oner$ |
$-{1\over N_c}$ | $0$
| $1-{{N_c-2}\over {2 m_f}}$ \cr
$X$ | ${\bf asym}$ | $\oner$ | $\oner$ | $\oner$ | ${2\over N_c}$ | $1$ |
$\half$ \cr
$\tilde X$ | $\overline{\bf sym}$ | $\oner$ | $\oner$ | $\oner$ |
$-{2\over N_c}$
| $-1$ | $\half$
\endtable

Setting $\mu$ to $1$ for convenience,
the superpotential takes the form
\eqn\esuper{
W = \Tr (X \tilde X)^2 + \hat Q \tilde X \hat Q~,
}
and truncates the chiral ring via the relations
\eqn\ringrel{
\tilde X (X \tilde X) =0, \quad 2 (X \tilde X X)_a^b + \hat Q_a \hat Q^b =0,
\quad \tilde X \hat Q =0~.
}
The chiral mesons are $M_0^{f\dot g} = Q^f \tilde Q^{\dot g}$,
${\tilde M}^{i \dot g} = \hat Q^i \tilde Q^{\dot g} $, $M_1^{f \dot
g} = Q^f  \tilde X X \tilde Q^{\dot g}$, $P^{fg}= Q^f \tilde X Q^g$
and $\tilde P^{\dot f \dot g} = \tilde Q^{\dot f} X \tilde Q^{\dot
g}$. The meson $P$ is symmetric in flavor indices and $\tilde P$ is
antisymmetric.

Note the matter content of this theory is similar to one of the
models considered in \intril,
where $SU(N_c)$ gauge theory with matter corresponding to a symmetric
and an antisymmetric multiplet, and $m_f+8$ fundamentals with
$SU(m_f+8)_L$ flavor symmetry and $m_f$ anti-fundamentals with
$SU(m_f)_R$ flavor symmetry. Our model may be obtained by adding the
$\hat Q$ dependent term of the superpotential \esuper\ to that of
\intril. Likewise, the model of \intril, may be obtained from our
model by adding mass terms of the form $\hat Q \tilde Q$, and
integrating out the massive fermions. This deformation will be
considered in more detail later in this section.

This theory has a non-anomalous $\IZ_{8m_f}$ discrete symmetry
generated by:
\eqn\edissym{
\eqalign{
(X, ~\tilde X) & \to \alpha^{2m_f} ~ (X, ~\tilde X) \cr
\hat Q &\to \alpha^{-m_f} \hat Q \cr
(Q,~ \tilde Q) & \to \alpha^{4-2N_c} (Q, ~\tilde Q) ~,\cr}
}
where $\alpha = e^{2\pi i/ (8m_f)}$.

\subsec{Magnetic Theory}

The theory has gauge group $SU(\tilde N_c)$ where $\tilde N_c = 3 m_f
+4 -N_c$ and a symmetric and an antisymmetric chiral multiplet together
with $m_f+8$ fundamental and $m_f$ anti-fundamental chiral multiplets,
and gauge singlet fields $M_0^{f \dot g}$, $\tilde M^{i \dot g}$,
$M_1^{f \dot g}$, $P^{fg}$ and $\tilde P^{\dot f \dot g}$.
The fields transform under the symmetries as:
\thicksize=1pt
\vskip12pt
\begintable
\tstrut |$ SU(\tilde N_c) $|$ SU(m_f)_L $|$ SU(m_f)_R $|$ SO(8)_L
$|$U(1)_B$|
$ U(1)_X $|$ U(1)_R$ \crthick
$\hat q $|$ \tilde \ncr $|$ \oner $|$ \oner $|$ \evr $|
$ {1\over\tilde N_c} $|
${{N_c-m_f}\over {2 \tilde N_c}} $|$ {3\over 4} $\cr
$q $|$ \tilde \ncr $|$ \overline \mfr $|$ \oner $|$ \oner $|
$ {1\over\tilde N_c} $|$ {{m_f+2}\over \tilde N_c} $|
$ 1-{{\tilde N_c-2}\over {2 m_f}}$
\cr
$\tilde q $|$ \overline {\tilde \ncr} $|$ \oner $|$  \overline\mfr $|
$\oner $|
$-{1\over \tilde N_c} $|$ -{{m_f+2}\over \tilde N_c} $|
$ 1-{{\tilde N_c-2}\over {2 m_f}} $\cr
$Y $|$ {\bf asym} $|$ \oner $|$ \oner $|$ \oner $|$ {2\over \tilde N_c}
$|
${{N_c-m_f}\over \tilde N_c} $|$ \half $\cr
$\tilde Y$|$ \overline{\bf sym} $|$ \oner $|$ \oner $|$ \oner $|
$ -{2\over\tilde N_c} $|$ -{{N_c-m_f}\over \tilde N_c} $|$ \half$ \cr
$M_0 $|$ \oner $|$ \mfr $|$ \mfr $|$ \oner $|$ 0 $|$ 0 $|
$ 2 -{{N_c-2}\over m_f}$
\cr
$\tilde M $|$ \oner $|$ \oner $|$ \mfr $|$ \evr $|$ 0 $|$ \half $|
${7\over 4} -{{N_c-2}\over 2m_f}$ \cr
$M_1 $|$ \oner $|$ \mfr $|$ \mfr $|$ \oner $|$ 0 $|$ 0 $|
$ 3 -{{N_c-2}\over m_f}$
\cr
$P $|$ \oner$|$ {\bf sym} $|$ \oner $|$ \oner $|$ 0 $|$ -1 $|
$ {5\over 2}-{{N_c-2}\over m_f}$ \cr
$\tilde P $|$ \oner$|$ \oner $|$ {\bf asym} $|$ \oner $|$ 0 $|$ 1 $|
${5\over 2} -{{N_c-2}\over m_f} $
\endtable

The superpotential takes the form:
\eqn\magpot{
W = \Tr (Y \tilde Y)^2 + \hat q \tilde Y \hat q + M_1 q \tilde q +
M_0 q (\tilde Y Y) \tilde q + \tilde M \hat q \tilde q + P q \tilde Y
q + \tilde P \tilde q Y \tilde q~.
}

The $\IZ_{8m_f}$ discrete symmetry on the magnetic fields is
generated by:
\eqn\mdissym{
\eqalign{
(Y, ~\tilde Y) & \to e^{i\pi B m_f}\alpha^{2m_f} ~ (Y, ~\tilde Y) \cr
\hat q &\to e^{i\pi B m_f}\alpha^{-m_f} \hat q \cr
(q,~ \tilde q) & \to e^{i\pi B m_f}\alpha^{4-2\tilde N_c} (q, ~\tilde q) ~,\cr}
}
where $B$ is the baryon number operator. This symmetry is
non-anomalous in the magnetic theory.

\subsec{'t Hooft Anomaly Matching}

As a check on the duality predicted by the brane configuration, we
have checked the
't Hooft anomaly matching conditions and found that they are
satisfied. The results are summarized in the following:
\eqn\anomatch{
\matrix{
U(1)_R & -{N_c^2 \over 2} -1\cr
U(1)_X & 3 N_c \cr
U(1)_B & 6 \cr
U(1)_R^3 & {1\over 8} (7 N_c^2 -N_c -8) - {1\over 4 m_f^2} (N_c-2)^3
N_c \cr
U(1)_X^3 & 0 \cr
U(1)_B^3 & 0 \cr
U(1)_X U(1)_B U(1)_R & -N_c -1\cr
U(1)_B^2 U(1)_R & -3 \cr
U(1)_B^2 U(1)_X & 0 \cr
U(1)_X^2 U(1)_B & 0 \cr
U(1)_X^2 U(1)_R & -{{N_c(N_c+1)}\over 2} \cr
U(1)_R^2 U(1)_B & 4 \cr
U(1)_R^2 U(1)_X & 0 \cr
SU(m_f)^3 & N_c d_3(m_f)\cr
SO(8)^3 & 0 \cr
SU(m_f)^2 U(1)_R & -{{N_c (N_c-2)}\over 4 m_f} \cr
SU(m_f)^2 U(1)_X & 0 \cr
SU(m_f)^2 U(1)_B & {1 \over 2} \cr
SO(8)^2 U(1)_R & - {N_c \over 4} \cr
SO(8)^2 U(1)_X & {N_c \over 2}  \cr
SO(8)^2 U(1)_B & 1 \cr}
}

\subsec{Superpotential Deformations}

Let us now consider deformations of the
superpotential of the electric theory and examine how these are
mapped
to the magnetic theory. It is a further check on the duality that the
resulting low-energy theories also form a dual pair.

If we deform the superpotential of the electric theory by a term of
the form $m Q^1 \tilde Q^1$, the low-energy theory takes the same
form
with $m_f \to m_f-1$. In the magnetic theory, a term $m M_0^{1,1}$ is
added to the superpotential. Taking the vevs of the mesons, $\hat
q$, and $q^{g}$, $\tilde q^{\dot g}$ (for $g,\dot g>1$) to vanish,
the F-term equations reduce to
\eqn\masseom{
\eqalign{
\tilde q^1 Y \tilde Y q^1 &= -m \cr
\tilde q^1 q^1&=0 \cr
q^1 \tilde Y q^1 &=0 \cr
\tilde q^1 Y \tilde q^1 &=0 \cr
Y \tilde Y Y &=0 \cr
\tilde Y Y \tilde Y &=0 ~,\cr}
}
and the D-terms take the form \dterms.
A solution of these equations can be written in the form
$Y_{12} =1$, $Y_{21} = -1$,
$\tilde Y_{23}=1$, $\tilde Y_{32} =1$,
$(q^1)_3 = -\sqrt{2}$, $(\tilde q^1)_1=\sqrt{2}$ and $\nu=0$, with other
entries
zero. Here we have scaled $m$ to $2$ for convenience. The rank of the
gauge group is reduced by 3, to $SU(3 m_f -N_c +1)$.
The massless matter consists of a symmetric tensor, an antisymmetric
tensor,
8 fundamental chiral multiplets, and $m_f-1$ flavors of fundamental
and anti-fundamental chiral multiplets. The low-energy magnetic
theory
is thus the dual of the low-energy electric theory.

The other deformation we consider is adding a mass term of the form
$m \hat Q^1 \tilde Q^1$ to the electric superpotential. On the
magnetic side this corresponds to adding $m \tilde M^{1,1}$ to the
superpotential. Taking the vevs of the mesons, $\hat
q^i$ (for $i>1$), $q^{g}$, $\tilde q^{\dot g}$ (for $g, \dot g>1$) to
vanish,
the F-term equations now become
\eqn\ftermeom{
\eqalign{
\tilde q^1 Y \tilde q^1 &=0 \cr
\tilde Y \hat q^1 &=0 \cr
\tilde q^1 \hat q^1 &= -m \cr
\hat q^1_a \hat q^{1b} + 2(Y \tilde Y Y)_a^b &=0\cr
\tilde Y Y \tilde Y &=0 ~.\cr}
}
Scaling $m$ to $2\sqrt{2}$, a solution to these equations takes the
form
$(\tilde q^1)_1 = -2$, $(\hat q^1)_1 =\sqrt{2}$,
$Y_{12} =1$, $Y_{21} = -1$, $\tilde Y_{22} = 1$ with other
components
zero. This solves the D-term equations. The rank of the gauge group
is
broken by 2 to $SU(3m_f -N_c +2)$. The massless matter consists of a
symmetric tensor, an antisymmetric tensor, $m_f+7$ fundamental chiral
multiplets and $m_f-1$ anti-fundamental chiral multiplets.

Giving mass to all 8 $\hat Q$'s in the electric theory leaves one
with
just the $\Tr (X \tilde X)^2$ term in the superpotential. This is
precisely one of the theories considered in \intril. On the magnetic
side we see the gauge group is reduced to $SU(3m_f -N_c - 12)$ with
$m_f$ fundamentals and $m_f-8$ anti-fundamentals, plus symmetric and
antisymmetric tensors, which is precisely the magnetic dual found in
\intril. We see therefore that our theory is related to that of
\intril\ by a deformation. Likewise one could add the
$\hat Q \tilde X \hat Q$ term to the superpotential of \intril, to
recover the theory considered here.

\subsec{Flat Directions}

In this subsection we consider the low-energy theories on the
baryonic
branches of our original theory. The electric and magnetic
descriptions of these branches will be shown to form dual pairs,
providing a further check on the duality of the original theory.

Consider the branch where the baryon $\tilde B_n = \tilde X^n \tilde
Q^{N_c-n} \tilde Q^{N_c-n}$ gets an expectation value. Here the color
indices are contracted with two $\epsilon$-tensors. In the magnetic
theory this is mapped to
$\tilde b_n = \tilde Y^{2m_f+4-n} \tilde q^{m_f-N_c+n} \tilde
q^{m_f-N_c+n}$.
This expectation value breaks the gauge group to $SO(n)$ on the
electric side and leaves us
with $2m_f$ massless fundamentals (the superpotential generates a
mass
term for the adjoint matter field which may be integrated out).
In addition, there are
a number of singlets that decouple.
On the magnetic side we likewise have the gauge group
broken to $SO(2m_f+4-n)$ with $2m_f$ fundamentals.
This is the magnetic dual of the electric
theory as expected \seiberg.

 From the brane point of view, the baryonic branch with $n=N_c$
corresponds to moving the central NS-fivebrane in the positive $x^7$
direction. The displacement of the central fivebrane in the $x^7$
direction should be identified with the would-be FI parameter $-\nu$
appearing in the D-term equations \dterms. From this equation it is
clear that $X$ ($\tilde X$) can get an
expectation value for $\nu$ positive (negative).
A positive displacement generates a mass for the fundamentals
coming from the
half D-sixbranes and the D-fourbranes intersect the orientifold sixplane on the
side where the gauge group is projected down to $SO(N_c)$. However,
the brane picture only gives one real parameter for this baryonic
branch. Presumably if this picture is lifted to M-theory, a more
careful analysis of the curve along the lines of \ooguri\ will
give a complete picture of the moduli space.

Now consider the baryonic branch where $B_n = X^n Q^{N_c-2n}$ gets an
expectation value. This breaks the gauge group to $Sp(n)$ and leaves
us with $2m_f+8$ massless fundamentals,
plus some singlets that decouple. The superpotential takes the form
$W= \Tr \hat Q^4/\lambda$, where $\lambda$ is proportional to the vev of $X$.
On the magnetic side
the baryon operator is mapped to
$b_n = Y^{m_f+2-n} q^{m_f-N_c+2n}$. This breaks the gauge group to
$Sp(m_f+2-n)$ and leaves us with $2m_f+8$ massless fundamentals plus
some singlets. This theory is the magnetic dual expected from
\seiberg,
with an additional deformation of the superpotential, which vanishes
in the limit $\lambda \to \infty$.

In terms of branes, for $n=N_c/2$ (with $N_c$ even), this corresponds
to moving the central NS-fivebrane in the negative $x^7$ direction.
The 8 special fundamentals remain massless on this branch, and the
gauge group is projected to $Sp(N_c/2)$ when the D-fourbranes
intersect this section of the O-sixplane. Again we only see one real
parameter of this branch in a simple way in the brane picture.

Finally, let us consider the baryonic branches where
$\hat B_{n,a} = X^n \hat Q^a Q^{N_c-2n-a}$ gets an expectation value.
The gauge group is broken to $Sp(n)$ as before, but the massless
matter content is different in this case. The F-terms give
\eqn\bfterms{
\eqalign{
\tilde X \hat Q &=0\cr
\tilde X X \tilde X &=0\cr
\hat Q_a \hat Q^b + 2 (X \tilde X X)_a^b &=0~,\cr
}}
which implies that both $X$ and $\tilde X$ receive a vev. Solving
these equations we find the massless matter is $2m_f+8-2a$
fundamentals of $Sp(n)$.
The baryon is mapped to
$\hat b_{n,a} = Y^{m_f+2-n-a}  \hat q^a q^{m_f-N_c+2n+a}$ on the
magnetic side. Likewise, this breaks the gauge group to
$Sp(m_f+2-n-a)$ with $2m_f+8-2a$ massless fundamentals. This is the
expected magnetic dual \seiberg.

\newsec{Conclusions and Outlook}

We investigated a brane configuration that genuinely gives chiral
matter
in its effective four-dimensional field theory description. The key
ingredient
is the orientifold sixplane that is divided in two by an
NS-fivebrane.
Using linking number arguments along the lines of \brodie\ we were
able to
obtain the dual model from brane moves. This is of course only a
first
step
towards gaining a deeper understanding of chiral $\CN=1$ theories
from
branes.

Still in our model there remain many unanswered questions. We
considered
the brane configuration only in Type \IIA\ string theory. In order to
obtain
more information about the strong coupling behavior one should lift
the
configuration to M-theory and derive a complex curve describing the
system.
The existence of a smooth curve seems plausible since the sixbrane
charge
does not jump at the location of the center NS-fivebrane thanks to
the
additional eight fundamentals $\hat Q$. The role these special fields
play
is
very interesting. From the field theory point of view there is
actually no
reason why one should couple precisely eight of the fundamentals to
the
symmetric tensor. Indeed we saw that our model can also be seen as a
deformation of the model in \intril\ and vice versa. A brane
configuration
giving rise to a coupling of less (or more)
than eight fundamentals to the symmetric tensor needs the
introduction of
additional D-eightbranes. This is because the sixbrane charge would jump
now
at the location of the center NS-fivebrane and this is only
consistent in
the
presence of D-eightbranes \hazaff.

Another generalization is to consider
also
configurations with coinciding fivebranes at the
positions of our A-, B- and C-fivebranes. These should give
field
theories with superpotentials of the form $\Tr (X\tilde X)^{2(k+1)}$.
Putting additional D-sixbranes on top of the orientifold presumably
gives
theories with couplings not only of fundamentals to the symmetric
tensor
but also of anti-fundamentals to the antisymmetric tensor. All these
points
deserve further investigation and
we plan to elaborate on these issues in future work \future.

\bigskip

\centerline{\bf Acknowledgments}

We wish to thank C. Gomez, A. Hanany, K. Intriligator and E.
Rabinovici for discussions.
We thank C.E.R.N. and I.F.T. at U.A.M. for hospitality during the
completion of this work. K.L. was supported by FWF Project Nr.
P10268-PHY.
E.L. was supported by Lise Meitner Fellowship M456-TPH.
D.L. was supported in part by DOE grant DE-FG0291ER40688-Task A.

\listrefs
\end